\documentclass[a4paper,journal]{IEEEtran}
\usepackage{amsmath,amsfonts}
\usepackage{algorithmic}
\usepackage{algorithm}
\usepackage{array}
\usepackage[caption=false,font=normalsize,labelfont=sf,textfont=sf]{subfig}
\usepackage{textcomp}
\usepackage{stfloats}
\usepackage{url}
\usepackage{verbatim}
\usepackage{graphicx}
\usepackage{cite}
\usepackage{multirow}
\usepackage{multicol}
\usepackage{booktabs}
\begin{document}

\title{Sequence-to-Sequence Neural Surrogate Modeling of Transient Responses in High-Speed Links with Variable Design Parameters}

\author{Julian Withöft, Shayan Ghanbari, Werner John, Emre Ecik, Ralf Brüning, Jürgen Götze
\thanks{This work has been submitted to the IEEE for possible publication. Copyright may be transferred without notice, after which this version may no longer be accessible.}%
\thanks{This work is funded as part of the research project KI4BoardNet in the funding programme MANNHEIM (BMBF) (Grant numbers 16ME0779/TUDO - 16ME0777/IDMT). The responsibility for this publication is held by the authors only.}%
\thanks{Julian Withöft, Shayan Ghanbari, Emre Ecik and Jürgen Götze are with the Information Processing Lab, Faculty for Electrical Engineering and Information Technology, TU Dortmund, Germany (e-mail: \{julian.withoeft, shayan.ghanbari, emre.ecik, juergen.goetze\}@tu-dortmund.de)}%
\thanks{Werner John is with Pyramide2525, Paderborn, Germany and the Information Processing Lab, Faculty for Electrical Engineering and Information Technology, TU Dortmund, Germany (e-mail: werner.john@pyramide2525.de)}%
\thanks{Ralf Brüning is with EMC Technology Center Paderborn, Zuken GmbH, Paderborn, Germany (e-mail: ralf.bruening@de.zuken.com)}%
}

\maketitle

\begin{abstract}
This paper presents sequence-to-sequence (seq2seq) neural surrogates for high-speed transient response modeling and signal integrity analysis in fixed and variable design parameter spaces. To identify the structural requirements for prediction accuracy, multiple neural architectures were evaluated with respect to different training strategies such as sample-based and unit interval-based, as well as input configurations such as TX-only and TX+RX. For fixed design parameters, feedforward neural networks achieve accuracy comparable to recurrent models while requiring lower computational cost, as the static mapping can be approximated by fully connected and non-sequential representations. Variable design parameters alter the electromagnetic characteristics of the interconnect, which creates parameter-dependent temporal dynamics that require sequence-aware architectures to resolve the evolving interactions within the transient response. The results based on a 14-dimensional variable design parameter scenario show that while autoregressive TX+RX feedback suffers from severe phase drift and error propagation under variable conditions, non-autoregressive TX-only inference leads to stability. To overcome the structural limitations of standard decoders, a recurrent Early Fusion LSTM is proposed, which injects design parameters at the initial encoding stage. This approach achieves superior prediction accuracy, with an RMSE of 31.0~mV and an $\mathbf{R^2}$ of 99.51\% for transient waveform prediction, an RMSE of 30.7~mV and an $\mathbf{R^2}$ of 99.57\% for eye-height prediction, and an RMSE of 19.5~ps and an $\mathbf{R^2}$ of 99.94\% for eye-width prediction. These results demonstrate that neural seq2seq surrogates are accurate for transient response modeling and signal integrity analysis in complex variable design parameter spaces.
\end{abstract}


\section{Introduction}
\IEEEPARstart{R}{ising} data rates and increasing printed circuit board (PCB) complexity have made signal integrity (SI) a critical aspect of any high-speed system design today. Accurate time-domain transient simulations are essential for the evaluation of SI metrics such as eye diagrams as they provide information on the full dynamic behavior of interconnects. However, these simulations are often computationally expensive, especially when long bit sequences are involved. To address these challenges, machine learning (ML) has emerged as a promising approach to accelerate existing simulation frameworks to replace the repetitive long bit response simulation with predictive surrogate models. These methods can significantly reduce simulation time at high accuracy, which enables faster design iterations and efficient exploration of large design spaces. This is particularly beneficial to reduce the number of design cycles required by the PCB developer to ensure SI. As a result, data-driven modeling is becoming an increasingly relevant tool in modern SI analysis.

The increasing demand for faster and more efficient SI analysis in high-speed digital systems has led to a notable increase in ML-based modeling approaches, especially for transient response prediction.

Early work in \cite{Nguyen2018,Nguyen2019} demonstrated that recurrent neural networks (RNNs) and long short-term memory (LSTM) models can learn the temporal behavior of high-speed links based on the example of a single transmission line with fixed geometric design parameters. These models predict the receiver (RX) voltage sample-by-sample autoregressively based on a lookback of past transmitter (TX) and RX values, using a teacher-forcing strategy during training. Accurate predictions were achieved for both NRZ and PAM-4 signaling.

Further research in \cite{Goay2021} combined convolutional layers with LSTMs to create a hybrid convolutional neural network (CNN)-LSTM approach, which enhances feature extraction and improves prediction accuracy. Their approach also utilized adaptive hyperparameter tuning to accelerate convergence of the models. More complex interconnect structures compared to the simple transmission line in \cite{Nguyen2018,Nguyen2019} were analyzed, but the design parameters remained fixed as before. In \cite{Goay2023}, temporal convolutional networks (TCNs) were proposed to replace recurrent architectures with dilated causal convolutions and residual blocks to achieve longer memory capacity with lower computational cost than LSTMs or CNN-LSTMs. The TCN showed faster convergence, reduced training data requirements, and superior accuracy for different high-speed channels. It also allowed models to utilize smaller lookback and larger lookahead windows for higher efficiency in fixed design parameter scenarios.

While the aforementioned works relied on RX-dependent training, TX-only approaches also emerged as an efficient alternative, because they eliminate the need for RX feedback and autoregressive (AR) rollout through prediction of channel responses solely from TX sequences at the input. In \cite{Chu2019,Chu2021}, this principle was demonstrated using fully connected neural networks to model RX responses from TX bit streams, while \cite{Luo2023} further advanced the approach with bidirectional LSTM (BiLSTM) architectures, which achieved superior performance in both eye metric reconstruction and waveform prediction. Additionally, in \cite{Chu2021,Luo2023}, validation with oscilloscope measurements was conducted, which confirmed strong agreement between predicted, simulated, and measured transient responses and eye diagrams.

Other improvements have been focused on training stability and accuracy at signal edges in fixed design parameter setups. Batch normalization was incorporated into RNNs in \cite{Faraji2022}, which helped to mitigate prediction errors at sharp transitions of the transient waveform, while also stabilizing and shortening training. In \cite{Ma2024,Ma2025}, a deep simple recurrent unit (SRU) model was proposed to accelerate training and improve signal prediction compared to gated recurrent unit (GRU), LSTM, and RNN models, especially around signal edges.

In \cite{Qiu2024a,Qiu2024b}, transfer learning was used for initialization of models for unseen interconnect setups based on a pre-trained deep SRU. This improved convergence and accuracy compared to Xavier initialization. The drawback of transfer learning is that it requires some retraining for each new configuration. Therefore, it does not provide a unified surrogate that generalizes on the whole continuous parameter space.

A different approach can be found in \cite{Li2023}, which uses a two-stage graph neural network (GNN)-RNN cascade to model both driver and channel behavior based on varying topologies. However, the evaluated channel conditions remain relatively slow, as the fastest rise and fall times are restricted to 2~ns, which limits the applicability to modern high-speed systems.

In summary, while these studies have clearly advanced the state of the art, they are confined to single-link modeling assuming fixed structural PCB configurations, rely on transfer learning to apply the model to different channel conditions, or allow for variable channel parameters and topologies, but operate under limited high-speed conditions. As a consequence, most methods lack the ability to generalize over a continuous range of physical design parameters, which is critical in the early design space exploration phase of high-speed systems. Moreover, while AR rollout performs well for fixed configurations, its prevalence in the literature introduces computational inefficiencies and susceptibility to error accumulation when scaled to variable design spaces. Finally, many studies directly accept the complexity of recurrent neural networks due to the temporal structure of the input and output data without previous analysis of simpler feedforward neural network structures.

To address these limitations, this work introduces a sequence-to-sequence (seq2seq) neural surrogate modeling framework for the prediction of transient responses in high-speed interconnects with variable design parameters. When design parameters such as transmission line lengths, termination resistors, or characteristic impedances are varied, they alter the electromagnetic characteristics of the channel. This creates complex parameter-dependent temporal dynamics that often lead to severe error accumulation in AR models. When formulating transient modeling as a seq2seq problem, the architecture decouples the feature extraction of the input history and design parameters in the encoder from the generation of the output sequence in the decoder. The separation of encoder and decoder allows for the combination of design parameters and the dynamic signal history, which results in a surrogate model that can learn the relationship between a high-dimensional continuous design space and the transient waveforms. The core contributions of this work are as follows:

\begin{itemize}
	\item \textbf{Evaluation of Seq2Seq Formulations:} A comparison of four modeling configurations is conducted with sample-based vs. unit interval (UI)-based data representations, and AR (TX+RX) vs. non-AR (TX-only) inference modes. The results reveal that while AR feedback is common in the literature for fixed design parameter scenarios, it is prone to error propagation and numerical instability in variable design parameter scenarios.
	\item \textbf{Architectural Comparison for Fixed vs. Variable Design Parameters:} For static and fixed-parameter setups, the results demonstrate that computationally lightweight fully connected networks are on par with complex recurrent models in terms of predictive accuracy. In contrast, it is shown that variable design parameter spaces require sequence-aware architectures to resolve shifting electromagnetic temporal dynamics.
	\item \textbf{Superior Accuracy via Early Fusion LSTMs:} In variable cases, the Early Fusion LSTM architecture, which injects the static physical design parameters into the initial sequence encoding stage, is more accurate compared to any intermediate fusion architecture that fuses sequence history with design parameters in the latent space, and even outperforms transformer architectures.
	\item \textbf{Continuous 14-Dimensional Design Space Generalization:} Integration of the non-AR seq2seq formulation with early parameter fusion leads to a universal LSTM surrogate capable of learning the relationship between a complex 14-dimensional continuous design parameter space and the output transient responses. It achieves an RMSE of 31.0~mV at an $R^2$ of 99.51\% for transient waveform prediction, 30.7~mV RMSE with 99.57\% $R^2$ for eye-height prediction, and 19.5~ps RMSE with 99.94\% $R^2$ for eye-width prediction on the test dataset.
\end{itemize}

\section{PCB Interconnect Structures}

Two PCB interconnect structures for transient response modeling were considered, including applications with fixed and variable design parameters.

\subsection{DDR3 DQ Point-to-Point Interconnect}

\begin{figure}[t!]
	\centering
	\includegraphics[width=8.5cm]{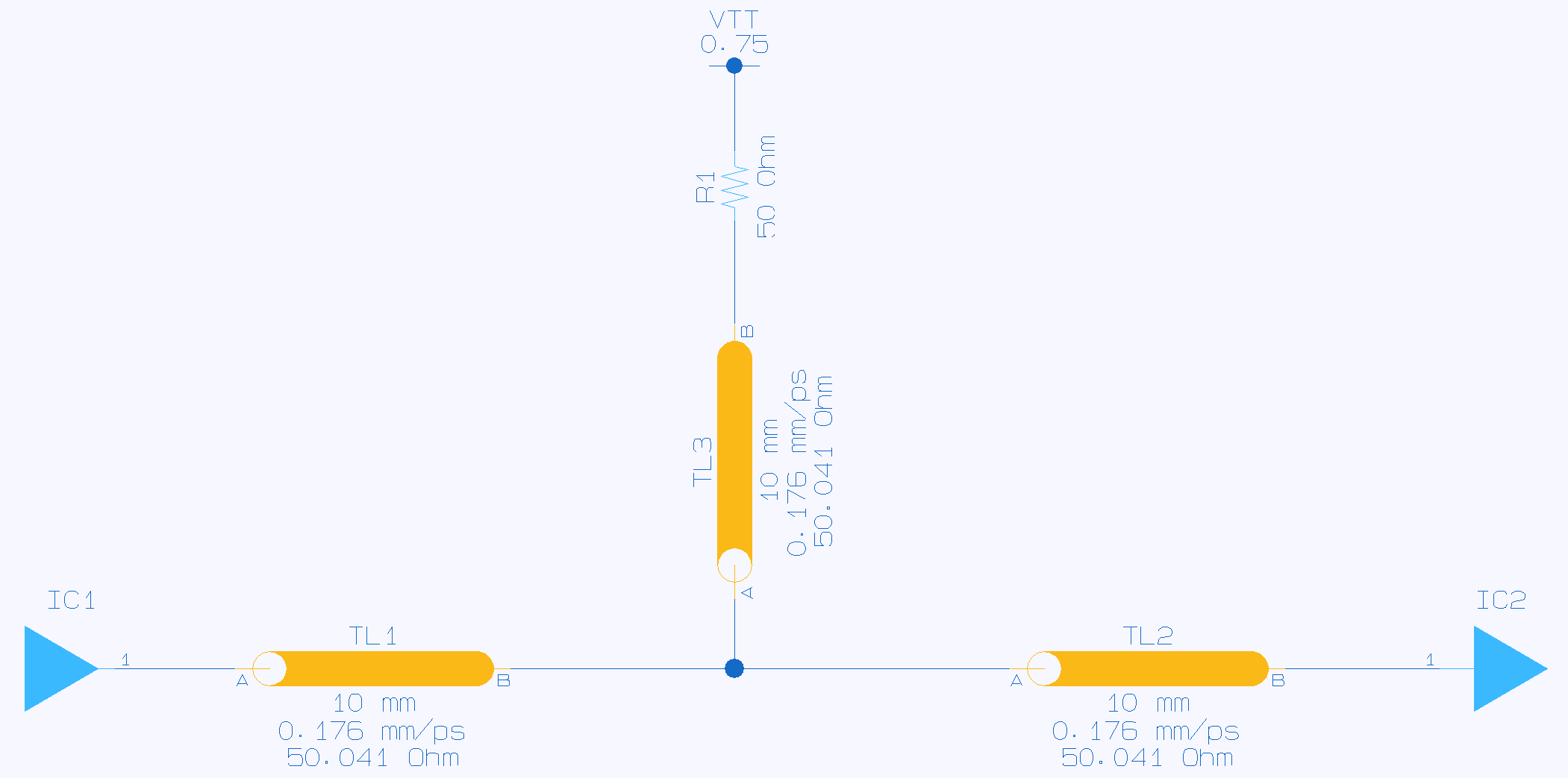}
	\caption{DDR3 DQ point-to-point interconnect \cite{Withoeft2025}.}
	\label{fig_ddr3_interconnect}
\end{figure}

Primarily, the DDR3 DQ point-to-point interconnect shown in Fig.~\ref{fig_ddr3_interconnect} was utilized \cite{Withoeft2025}. This is a memory system with a single TX and RX, and three transmission line segments terminated by resistor $R_1$ at the VTT midpoint supply voltage (VDDQ/2). The TX includes a variable internal resistance $R_{\text{in,tx}}$ and capacitance $C_{\text{in,tx}}$, while the RX consists of a variable input capacitance $C_{\text{in,rx}}$ and a high-impedance 1~M$\Omega$ input resistance. This structure is evaluated based on transient simulations in LTspice \cite{LTspice}. To ensure that the surrogate models learn dynamic and generalized signaling behaviors, the parameter space summarized in Table~\ref{tab_ddr3_parameters} also varied the stimulus parameters, such as the clock frequency $f_{\text{clock}}$, supply voltage VDDQ, as well as rise and fall times $t_r$/$t_f$, in addition to these PCB design parameters.

The data was generated using two scenarios, namely the fixed and the variable design parameter cases. For the fixed design parameter scenario, 7 random channel configurations within the ranges of Table~\ref{tab_ddr3_parameters} were simulated. Each fixed case was stimulated with a pseudo-random bit sequence (PRBS)-15 stimulus to generate a 32,768-bit response, which was then chronologically partitioned based on a 70-15-15\% train-validation-test split. The variable design parameter scenario used Latin Hypercube Sampling (LHS) to cover the entire design space of Table~\ref{tab_ddr3_parameters}. For this case, a PRBS-10 stimulus was applied to each unique configuration to generate a large dataset with 40,000 training, 10,000 validation, and 10,000 test samples. This leads to an aggregate bit volume of approximately 40 million training, 10 million test, and 10 million validation bits and results in a dense design parameter space coverage for surrogate model training without requiring extended bit sequences per sample.

\begin{table}[t!]
	\centering
	\caption{PCB and IC parameters and their ranges for the DDR3 DQ point-to-point interconnect \cite{Withoeft2025}}
	\label{tab_ddr3_parameters}
	\begin{tabular}{lcc}
		\toprule
		\textbf{Parameter} & \textbf{Min.} & \textbf{Max.} \\
		\midrule
		Clock Frequency $f_{\text{clock}}$ & 100 MHz & 800 MHz \\
		Supply Voltage (VDDQ) & 1.2 V & 2.5 V \\
		Rise and Fall Time $t_r = t_f$ & 100 ps & 500 ps \\
		TX Resistor $R_{\text{in,tx}}$ & 10 $\Omega$ & 50 $\Omega$ \\
		TX Capacitance $C_{\text{in,tx}}$ & 0.1 pF & 7 pF \\
		RX Capacitance $C_{\text{in,rx}}$ & 0.1 pF & 7 pF \\
		Transmission Line Length TL1 & 1 mm & 200 mm \\
		Transmission Line Length TL2 & 1 mm & 25 mm \\
		Transmission Line Length TL3 & 1 mm & 200 mm \\
		Characteristic Impedance $Z_{0,\text{TL1}}$ & 30 $\Omega$ & 90 $\Omega$ \\
		Characteristic Impedance $Z_{0,\text{TL2}}$ & 30 $\Omega$ & 90 $\Omega$ \\
		Characteristic Impedance $Z_{0,\text{TL3}}$ & 30 $\Omega$ & 90 $\Omega$ \\
		Termination Resistor $R_1$ & 1 $\Omega$ & 200 $\Omega$ \\
		Dielectric Constant $\epsilon_{eff}$ & 3 & 5 \\
		\bottomrule
	\end{tabular}
\end{table}

\subsection{DDR4 CK Daisy Chain Interconnect}

\begin{figure}[t!]
	\centering
	\includegraphics[width=8.5cm]{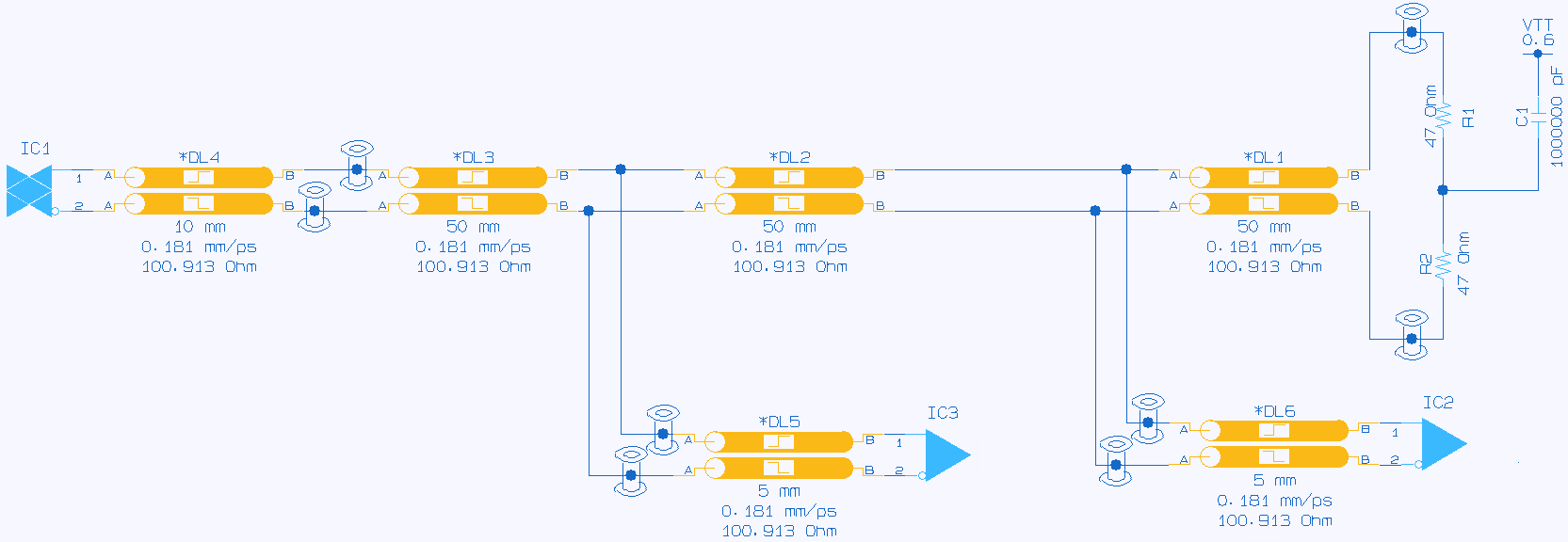}
	\caption{DDR4 CK daisy chain interconnect with transient predictions evaluated at the first RX IC3.}
	\label{fig_ddr4_interconnect}
\end{figure}

To further evaluate the fixed modeling scenario, the DDR4 CK daisy chain interconnect shown in Fig.~\ref{fig_ddr4_interconnect} was also analyzed. This topology has one TX and two RX, which are modeled as I/O buffer information specification (IBIS) models, specifically a Samsung K4A4G045WD model for the two RX and a Micron MT40A2G4 model for the TX. Transient predictions were evaluated at the first RX IC3. Fig.~\ref{fig_ddr4_interconnect} shows the interconnect structure with multiple coupled differential-pair microstrip transmission lines of different lengths and approximately 100~$\Omega$ differential impedance as well as an RC termination branch with two 47~$\Omega$ resistors and 1~µF capacitance and several vias along the signal path. The clock frequency was set to 1600~MHz in accordance with the DDR4-3200 speedgrade. Simulations were performed using the Zuken eCADSTAR SI software \cite{eCADSTAR}. Consistent with the previous fixed-parameter scenario, a PRBS-15 stimulus and the identical 70-15-15\% data split were utilized.

\section{Neural Networks}

Neural networks are universal function approximators that learn nonlinear input-output relationships \cite{Hornik1989}. They are based on fully connected layers, which compute
\begin{equation}
	\mathbf{a}^{[l]} = \phi\!\left(\mathbf{W}^{[l]}\mathbf{a}^{[l-1]} + \mathbf{b}^{[l]}\right),
\end{equation}
where $\mathbf{W}^{[l]}$ and $\mathbf{b}^{[l]}$ are learnable weights and biases, and $\phi$ is a nonlinear activation function. The weights and biases are optimized iteratively via backpropagation using gradient-based optimizers such as Adam \cite{Kingma2014}. When all neurons between adjacent layers are connected, the network is referred to as a dense or fully connected network. Dense networks are simple from an architectural point of view, but act as an important baseline because their capacity to approximate arbitrary functions makes them competitive for many problem formulations \cite{Goodfellow2016}.

Many learning problems, however, include data with temporal structure, which motivates architectures that model these sequential dependencies. RNNs achieve exactly this using a hidden state $\mathbf{h}_t$ that is updated at each time step based on the current input and the previous state \cite{Elman1990}. At each time step $t$, the hidden state is updated as
\begin{equation}
	\mathbf{h}_t = \phi\!\left(\mathbf{W}_h\mathbf{h}_{t-1} + \mathbf{W}_x\mathbf{x}_t + \mathbf{b}\right),
\end{equation}
where $\mathbf{W}_h$ and $\mathbf{W}_x$ are learnable weight matrices and $\mathbf{x}_t$ is the input at time $t$. RNNs are architecturally simple, but cannot retain information over long sequences due to vanishing gradients during backpropagation through time \cite{Bengio1994}. LSTM networks \cite{Hochreiter1997} address this using a cell state $\mathbf{c}_t$ and three multiplicative gates, namely the input, forget, and output gates, that regulate the flow of information. The additive nature of the cell state update provides a clear gradient path, which enables the network to learn dependencies over longer time horizons. BiLSTM networks \cite{Schuster1997} process the sequence in both the forward and backward directions and concatenate the resulting hidden states to analyze the context from both past and future at every position. Several variants have been proposed to reduce the computational complexity of LSTMs in terms of their four interacting gate computations and separate cell state carried at each time step. The GRU \cite{Chung2014} merges the cell and hidden state and reduces this to two gates, while the SRU \cite{Lei2018} further parallelizes element-wise state updates for significantly faster training and inference. Both achieve comparable accuracy to LSTMs on many tasks while reducing model complexity, but neither consistently beats LSTMs in terms of accuracy.

\begin{figure*}[t!]
	\centering
	\includegraphics[width=\textwidth]{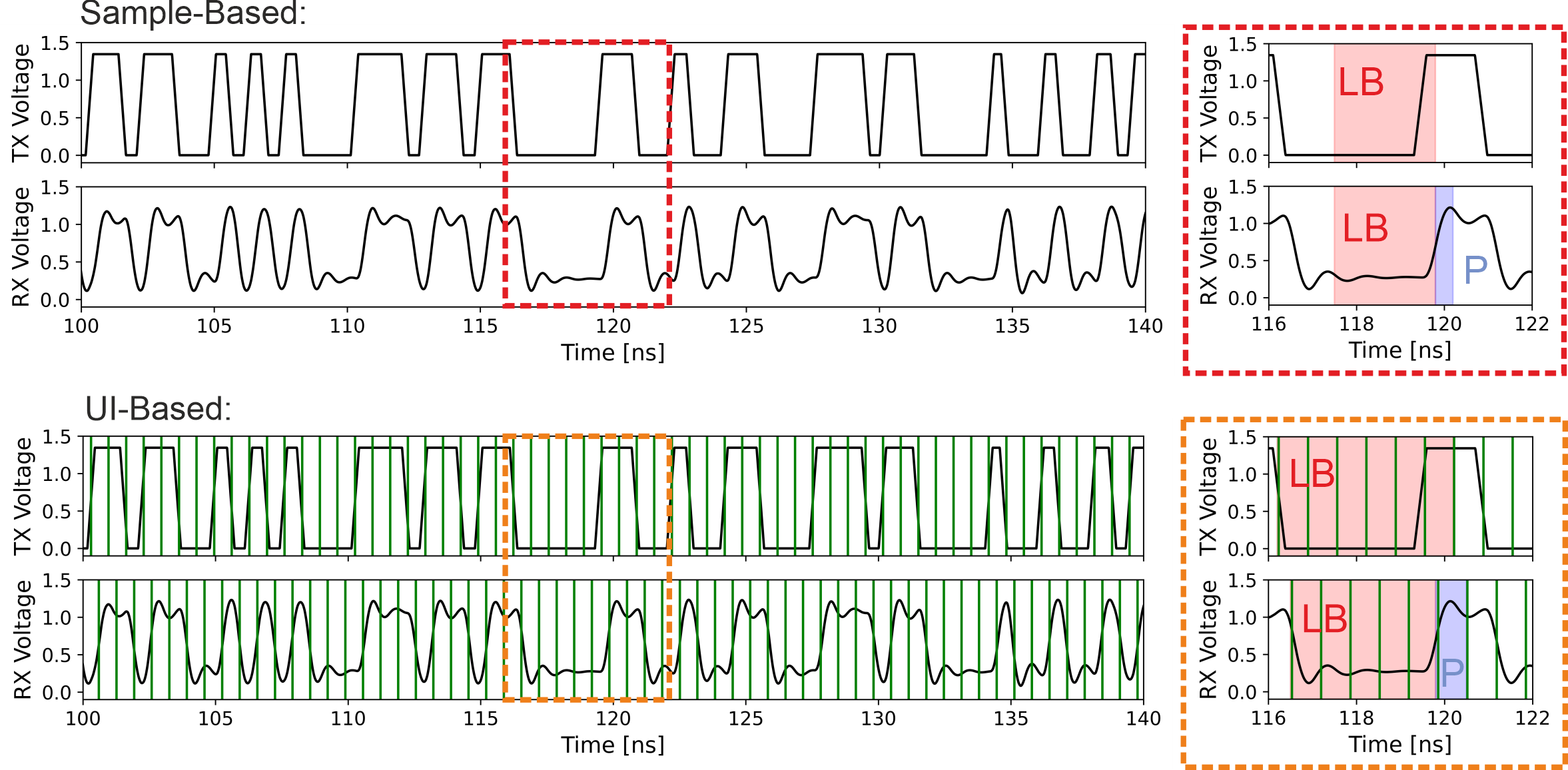}
	\caption{Illustration of sample-based (top) and UI-based (bottom) model input formulations. The lookback (LB) window ($W_{\mathrm{LB}}$) provides causal TX and RX context, while the prediction window (P) defines the model output target.}
	\label{fig_data_flow}
\end{figure*}

The Transformer architecture \cite{Vaswani2017} takes a completely different approach and replaces recurrence with self-attention. Every position in the sequence attends to all others simultaneously:
\begin{equation}
	\text{Attention}(\mathbf{Q},\mathbf{K},\mathbf{V}) = \text{softmax}\!\left(\frac{\mathbf{Q}\mathbf{K}^T}{\sqrt{d_{\text{head}}}}\right)\mathbf{V},
\end{equation}
where queries $\mathbf{Q}$, keys $\mathbf{K}$, and values $\mathbf{V}$ are linear projections of the input, and $\sqrt{d_{\text{head}}}$ is a scaling factor to prevent the dot products from becoming too large. Multiple attention heads are used in parallel on different learned projections to analyze dependencies based on different representations. This large receptive field makes the Transformer well-suited to sequence-related tasks where long-range dependencies are critical.

\section{Transient Modeling with Neural Networks}

At its core, transient response modeling with neural networks can be formulated as a sequence-to-sequence (seq2seq) modeling task \cite{Sutskever2014}. Given a lookback window of causal TX signals and optionally RX signals, the model predicts the next RX output. The formulation differs along two dimensions, namely the input representation (sample-based vs. UI-based) and the inference mode (AR TX+RX vs. non-AR TX-only), which leads to four modeling configurations as illustrated in Fig.~\ref{fig_data_flow}.

In the sample-based formulation, at discrete time index $k$, the TX and RX lookback vectors of $W_{\mathrm{LB}}$ raw time-domain samples are
\begin{align}
	\mathbf{x}^{\text{TX}}_k &= \left[v^{\text{TX}}_{k-W_{\mathrm{LB}}}, \ldots, v^{\text{TX}}_{k-1}\right]^T \in \mathbb{R}^{W_{\mathrm{LB}}}, \\
	\mathbf{x}^{\text{RX}}_k &= \left[v^{\text{RX}}_{k-W_{\mathrm{LB}}}, \ldots, v^{\text{RX}}_{k-1}\right]^T \in \mathbb{R}^{W_{\mathrm{LB}}},
\end{align}
and the model predicts the subsequent $P$ output samples $\hat{\mathbf{y}}_k = \left[v^{\text{RX}}_{k}, \ldots, v^{\text{RX}}_{k+P-1}\right]^T \in \mathbb{R}^{P}$. During inference, the temporal window advances to $k+P$ to generate the next output. This formulation and multi-step rolling forecast approach during inference are consistent with the prediction methodologies adopted in the literature \cite{Goay2021,Goay2023}. It should be noted that while inference advances by $P$ steps, the generation of training datasets typically utilizes overlapping windows to maximize coverage and data volume.

In the UI-based formulation, $W_{\mathrm{LB}}$ and $P$ are defined in units of UIs. Let $n_s = f_s \cdot T_{\text{UI}}$ denote the number of samples per UI. The TX lookback input includes the current UI being predicted in addition to the $W_{\mathrm{LB}}$ previous UIs, because UI-boundary alignment removes the TX-to-RX delay offset:
\begin{align}
	\mathbf{x}^{\text{TX}}_k &= \left[v^{\text{TX}}_{k - W_{\mathrm{LB}} \cdot n_s}, \ldots, v^{\text{TX}}_{k + P \cdot n_s - 1}\right]^T \in \mathbb{R}^{(W_{\mathrm{LB}}+P) \cdot n_s}, \\
	\mathbf{x}^{\text{RX}}_k &= \left[v^{\text{RX}}_{k - W_{\mathrm{LB}} \cdot n_s}, \ldots, v^{\text{RX}}_{k-1}\right]^T \in \mathbb{R}^{W_{\mathrm{LB}} \cdot n_s},
\end{align}
where the TX vector includes $(W_{\mathrm{LB}}+P) \cdot n_s$ samples covering both past and current TX, while the RX lookback includes only the past $W_{\mathrm{LB}} \cdot n_s$ samples. The prediction target is $\hat{\mathbf{y}}_k = \left[v^{\text{RX}}_{k}, \ldots, v^{\text{RX}}_{k+P \cdot n_s - 1}\right]^T \in \mathbb{R}^{P \cdot n_s}$. It should be noted that the UI processing needs reliable switching thresholds to function. Therefore, degraded channels in the variable design parameter space that lack these thresholds were excluded from this specific formulation, which results in a reduced UI-based dataset with 13,130 training, 3,375 validation, and 3,310 test samples.

To unify both the sample-based and UI-based formulations using a general surrogate framework, let $L$ denote the number of predicted samples per inference step, i.e., $L=P$ for the sample-based approach and $L=P \cdot n_s$ for the UI-based approach. The surrogate prediction for the TX+RX case at time index $k$, which is conditioned on the design parameters $\mathbf{p}$, is
\begin{equation}
	\hat{\mathbf{y}}_k = \mathcal{F}\!\left(\mathbf{x}^{\text{TX}}_k, \mathbf{x}^{\text{RX}}_k, \mathbf{p}; \boldsymbol{\theta}\right),
\end{equation}
where $\boldsymbol{\theta}$ denotes the learnable parameters of the neural network. In the fixed design parameter scenario, the relationship is reduced to $\mathcal{F}\!\left(\mathbf{x}^{\text{TX}}_k, \mathbf{x}^{\text{RX}}_k; \boldsymbol{\theta}\right)$ because the design parameters $\mathbf{p}$ can be omitted. In the AR TX+RX mode, the predicted output $\hat{\mathbf{y}}_k \in \mathbb{R}^L$ is fed back to update the RX lookback for the next inference step at time $k+L$:
\begin{equation}
	\mathbf{x}^{\text{RX}}_{k+L} = \left[\mathbf{x}^{\text{RX}}_{k}\!\left[L{:}\right],\, \hat{\mathbf{y}}_k\right]^T.
\end{equation}
Here, $\mathbf{x}^{\text{RX}}_{k}[L{:}]$ is the current RX lookback vector with the oldest $L$ entries discarded. The complete causal context is utilized, but there is also a risk of error accumulation over long prediction horizons.

In the TX-only mode, the RX lookback is omitted:
\begin{equation}
	\hat{\mathbf{y}}_k = \mathcal{F}\!\left(\mathbf{x}^{\text{TX}}_k, \mathbf{p}; \boldsymbol{\theta}\right).
\end{equation}
This eliminates the AR error propagation risk at the cost of reduced input information. All four combinations of input representation and inference mode are investigated in this work.

\begin{figure}[t!]
	\centering
	\includegraphics[width=8.5cm]{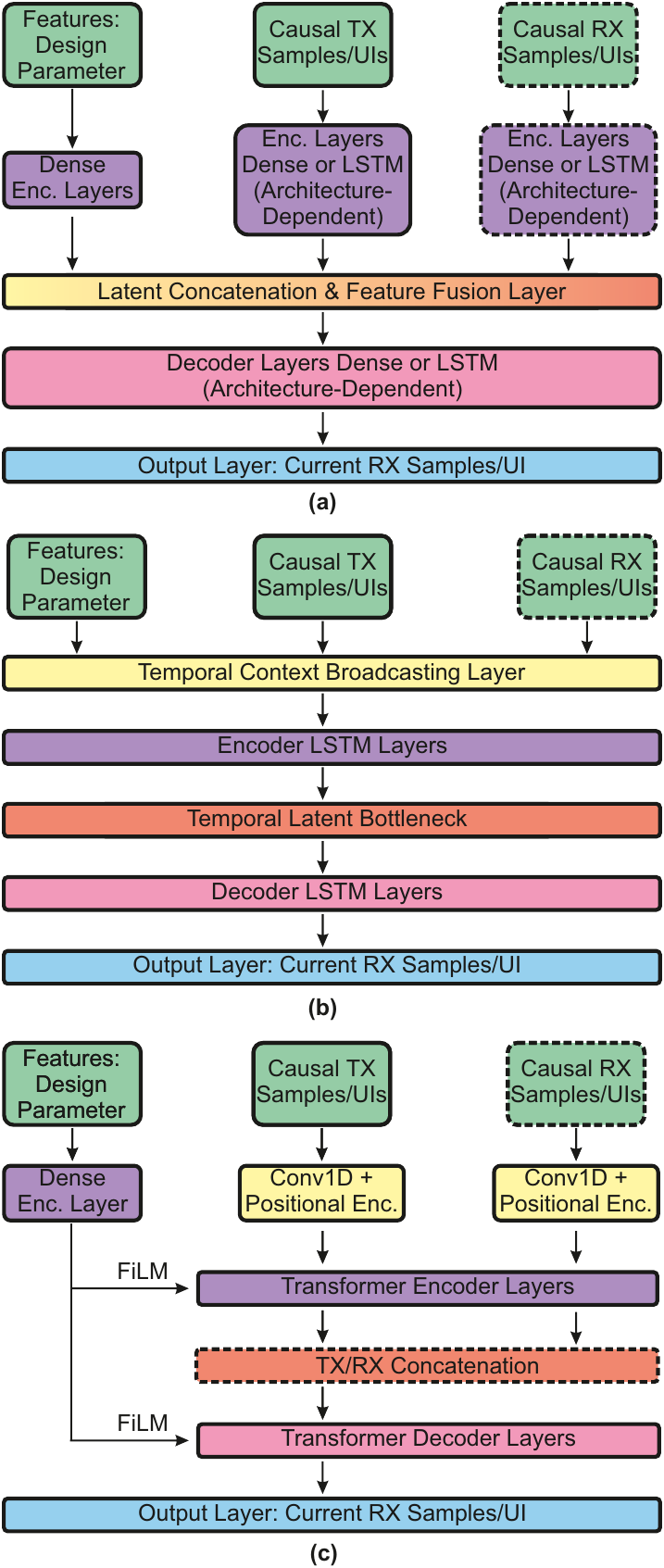}
	\caption{Illustration of intermediate fusion architectures (a), early fusion LSTM architecture (b), and transformer architecture (c).}
	\label{fig_ml_architectures}
\end{figure}

The seq2seq formulation of transient response modeling calls for encoder-decoder architectures \cite{Sutskever2014}, in which an encoder compresses the input lookback into a latent representation and a decoder generates the predicted output sequence from this representation. This structure separates the representation of the input sequences from the generation of the output. This is especially beneficial when design parameters have to be encoded with the signal history, as in the variable design parameter scenario. Three encoder-decoder architecture families are investigated in this work, as illustrated in Fig.~\ref{fig_ml_architectures}.

In the intermediate fusion architecture, Fig.~\ref{fig_ml_architectures} (a), the three inputs in form of design parameters, TX lookback, and RX lookback are encoded by separate branches and fused at a shared latent layer. Design parameters $\mathbf{p} \in \mathbb{R}^{N_p}$ are encoded by dense layers into a parameter embedding $\mathbf{z}_p \in \mathbb{R}^{d}$, while TX and RX lookback sequences are each passed through encoder layers, chosen as either dense or LSTM depending on the variant, which results in sequence embeddings $\mathbf{z}_{\mathrm{TX}}, \mathbf{z}_{\mathrm{RX}} \in \mathbb{R}^{d}$. These three embeddings are concatenated and provided to a feature fusion layer, followed by the decoder layers that produce the output sequence $\hat{\mathbf{y}} \in \mathbb{R}^L$. This leads to four main variants for the variable-parameter case, namely Dense-Dense, Dense-LSTM, LSTM-Dense, and LSTM-LSTM, where the notation refers to the encoder and decoder layer types, respectively. In the fixed design parameter experiments, the design-parameter branch is omitted. Therefore, Fig.~\ref{fig_ml_architectures} (a) is reduced to a pure sequence encoder-decoder with the same four combinations. For additional investigation, BiLSTM layers are also considered in the form of Dense-BiLSTM and BiLSTM-BiLSTM for the variable design scenario. Because this work aims to analyze the maximum predictive accuracy for transient modeling, the LSTM was selected as the recurrent baseline over lighter variants such as the GRU or SRU.

\begin{table*}[t!]
	\centering
	\caption{Fixed case test RMSE (mV) and $R^2$ (\%) for different training strategies, model architectures and interconnect scenarios}
	\label{tab_fixed_performance}
	\begin{tabular*}{\textwidth}{@{\extracolsep{\fill}} ll *{8}{c} @{}}
		\toprule
		\multirow{3}{*}{\textbf{Strategy}} & \multirow{3}{*}{\textbf{Model}} & \multicolumn{4}{c}{\textbf{DDR3 DQ P2P (Averaged over 7 runs)}} & \multicolumn{4}{c}{\textbf{DDR4 CK Daisy Chain}} \\
		\cmidrule(lr){3-6} \cmidrule(lr){7-10} 
		& & \multicolumn{2}{c}{\textbf{TX only}} & \multicolumn{2}{c}{\textbf{TX+RX (AR)}} & \multicolumn{2}{c}{\textbf{TX only}} & \multicolumn{2}{c}{\textbf{TX+RX (AR)}} \\
		\cmidrule(lr){3-4} \cmidrule(lr){5-6} \cmidrule(lr){7-8} \cmidrule(lr){9-10}
		& & \textbf{RMSE} & $\mathbf{R^2}$ & \textbf{RMSE} & $\mathbf{R^2}$ & \textbf{RMSE} & $\mathbf{R^2}$ & \textbf{RMSE} & $\mathbf{R^2}$ \\
		\midrule
		\multirow{4}{*}{\textbf{Sample-based}} 
		& Dense-Dense & 16.5 & 99.83 & 13.2 & 99.87 & 8.3 & 99.99 & 29.3 & 99.86 \\
		& Dense-LSTM & 16.7 & 99.81 & 12.8 & 99.87 & 9.4 & 99.99 & 26.3 & 99.88 \\
		& LSTM-Dense & 13.0 & 99.89 & 13.2 & 99.88 & 14.1 & 99.97 & 23.2 & 99.91 \\
		& LSTM-LSTM & 14.5 & 99.85 & 14.9 & 99.87 & 21.2 & 99.24 & 21.6 & 99.92 \\
		\midrule
		\multirow{4}{*}{\textbf{UI-based}} 
		& Dense-Dense & 13.1 & 99.95 & 36.6 & 99.26 & 16.8 & 99.95 & 8.4 & 99.99 \\
		& Dense-LSTM & 15.8 & 99.93 & 52.9 & 99.03 & 18.5 & 99.94 & 9.8 & 99.98 \\
		& LSTM-Dense & 11.1 & 99.96 & 32.9 & 99.36 & 16.2 & 99.96 & 2.9 & 99.99 \\
		& LSTM-LSTM & 11.2 & 99.96 & 36.0 & 99.57 & 16.0 & 99.96 & 2.7 & 99.99 \\
		\bottomrule
	\end{tabular*}
\end{table*}

In the Early Fusion LSTM architecture, Fig.~\ref{fig_ml_architectures} (b), the design parameters $\mathbf{p}$ are injected into the temporal input sequence before any recurrent processing. A temporal context broadcasting layer repeats $\mathbf{p}$ across the time dimension and concatenates it with the sequences, which results in a parameter-conditioned input tensor that is processed by the LSTM encoder layers. The final encoder state is compressed into a temporal latent bottleneck, which is then utilized by the LSTM decoder layers to create the output sequence. Early fusion encourages the network to integrate parameter information at every time step of the encoder, in contrast to intermediate fusion where parameter and sequence features are processed individually and are then only merged later in the latent space.

The Transformer architecture, Fig.~\ref{fig_ml_architectures} (c), replaces recurrent sequence processing with self-attention in the encoder and cross-attention in the decoder \cite{Vaswani2017}. TX and RX sequences first pass through one-dimensional convolutional layers to compress the long input sequences into tokens to reduce the computational complexity of self-attention. Sinusoidal positional encodings are then added to the tokens to inject absolute positional information \cite{Vaswani2017}. The design parameters $\mathbf{p}$ are projected through dense layers to generate feature-wise linear modulation (FiLM) parameters \cite{Perez2018}. The FiLM-conditioned encoder layers with self-attention process both TX and RX token sequences independently. TX and RX token sequences are concatenated along the token dimension after the encoder. The FiLM-conditioned decoder layers use cross-attention to process the concatenated TX+RX representations. Finally, a time-distributed linear head generates the output sequence $\hat{\mathbf{y}}$.

Prior to training, the design parameters $\mathbf{p}$ were normalized. All models utilized a hidden layer width of 256 neurons and were trained using the Adam optimizer \cite{Kingma2014} with a learning rate of $10^{-4}$, mean squared error (MSE) loss, a batch size of 128, and early stopping with a patience of 50 epochs. The fixed case used the full training set per epoch, while the variable case utilized 1000 batches per epoch for both training and validation with random design parameter and temporal window sampling within each batch.

\section{Transient Modeling with Fixed Design Parameters}

Initially, transient modeling with fixed design parameters was analyzed. In these cases, the interconnect structure is invariant, which eliminates the need for a conditioning design parameter vector $\mathbf{p}$. Therefore, the models were trained only on historical TX sequences and, in the AR case, causal RX sequences to predict the subsequent RX output, as shown in Fig.~\ref{fig_data_flow}.

All experiments utilized a sampling resolution of $n_s = 64$ samples per UI. The optimal temporal lookback window $W_{\mathrm{LB}}$ was different depending on the data representation and inference strategy. For the UI-based formulation, $W_{\mathrm{LB}\text{,UI}} = 16$ UIs was selected for both TX-only and TX+RX cases because the UI-based processing removes the propagation delay. For the sample-based strategy, $W_{\mathrm{LB}\text{,Sample,TX+RX}} = 200$ samples was optimal as the state context that was given by the RX feedback can be utilized for prediction. The sample-based TX-only configuration required an extended $W_{\mathrm{LB}\text{,Sample,TX-only}} = 2000$ samples to provide more detailed information on the propagation delay and extended channel memory without the causal RX context. To ensure maximum predictive accuracy for each configuration, these lookback values were analyzed with preliminary experiments and sensitivity analyses.

\begin{figure}[t!]
	\centering
	\includegraphics[width=8.5cm]{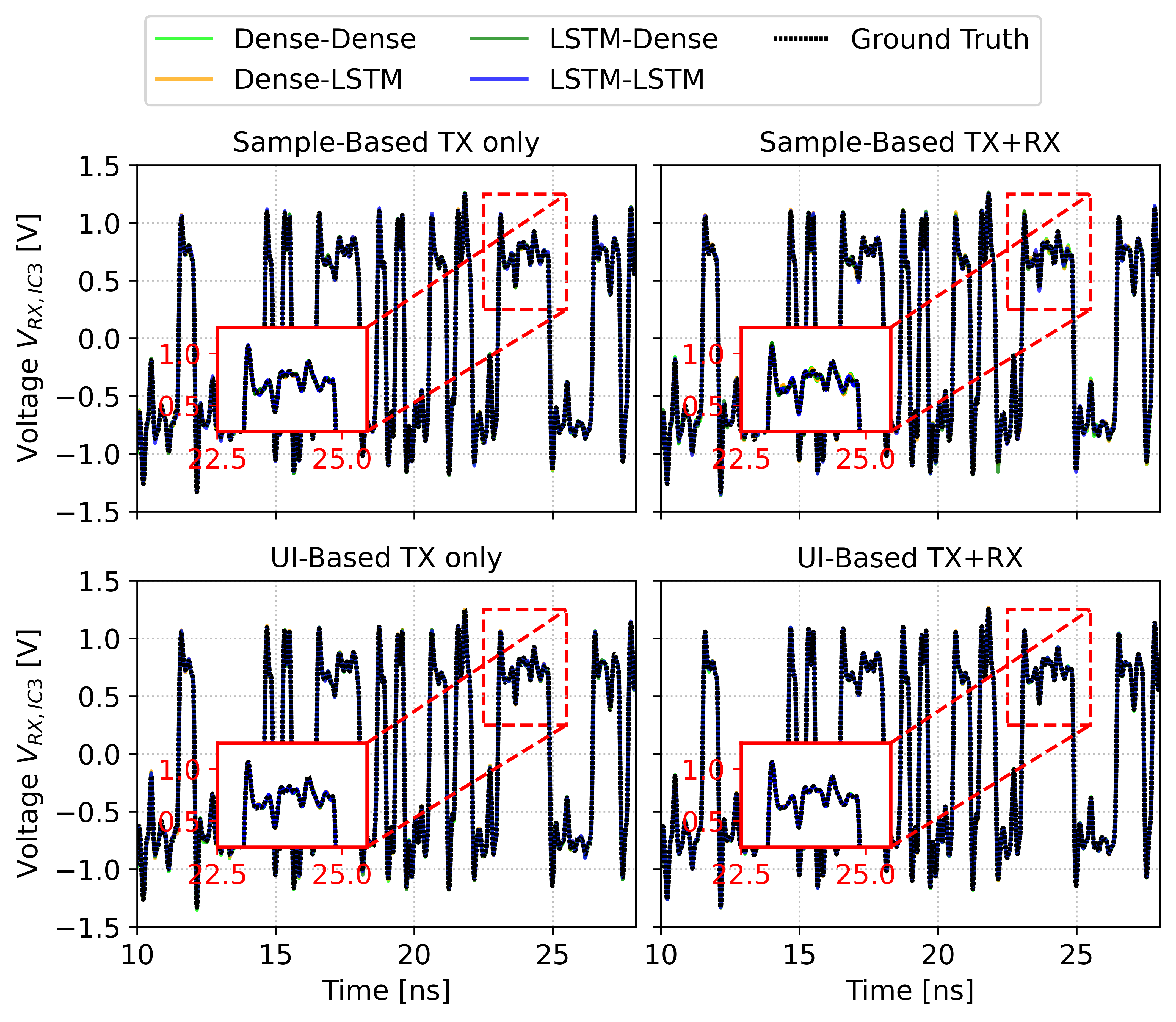}
	\caption{Transient response prediction examples for the DDR4 CK daisy chain interconnect.}
	\label{fig_ddr4_tran_response}
\end{figure}

\begin{figure}[b!]
	\centering
	\includegraphics[width=8.5cm]{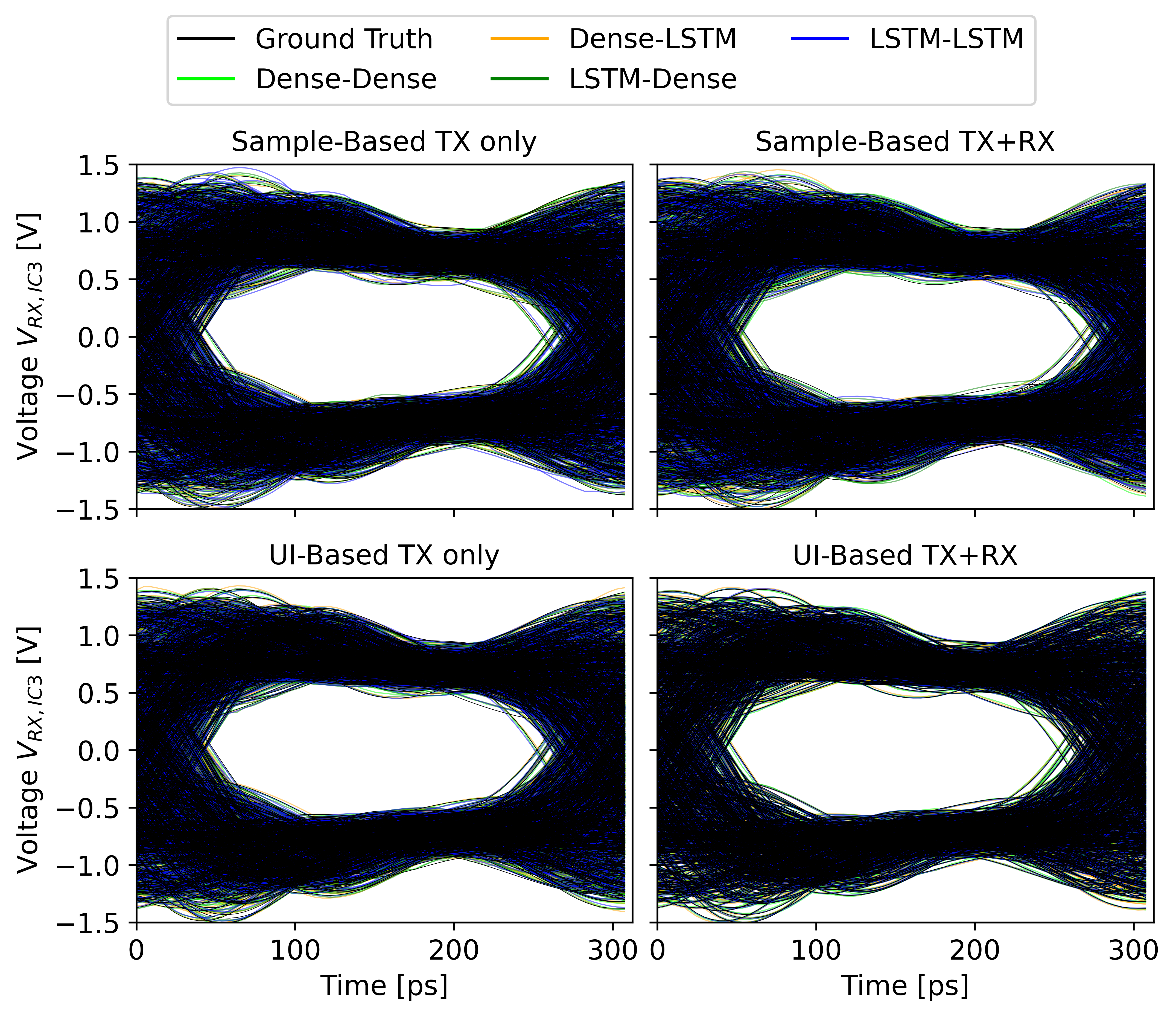}
	\caption{Eye prediction examples for the DDR4 CK daisy chain interconnect.}
	\label{fig_ddr4_eye_response}
\end{figure}

The transient prediction accuracy for the fixed design parameter cases based on the two considered interconnect structures is summarized in Table~\ref{tab_fixed_performance}. This indicates that all evaluated model architectures achieve high prediction accuracy for all of the considered training strategies, as the $R^2$-scores are above $99\%$ and the root mean squared error (RMSE) values are relatively small. The overall variance in RMSE is moderate, and there is no setting where one model consistently outperforms the others for all of the considered scenarios. Very importantly, the Dense-Dense architecture achieves predictive accuracy that is comparable to hybrid and fully recurrent configurations such as Dense-LSTM, LSTM-Dense, and LSTM-LSTM models. This demonstrates that all of the considered architectures model the underlying signal behavior with high accuracy for both of the investigated interconnect structures.

The predicted transient responses are compared to the ground-truth waveforms visually in Fig.~\ref{fig_ddr4_tran_response} based on the representative DDR4-3200 CK daisy chain interconnect for the four considered model architectures and the different training configurations, including sample-based and UI-based strategies as well as TX-only and TX+RX inputs. All models closely follow the reference waveform during the entire time window and even in the zoomed regions highlighting fine waveform details, where only minor deviations between the predictions and the ground truth can be observed.

To further evaluate the SI characteristics of the predicted transient responses, the corresponding eye diagrams are shown in Fig.~\ref{fig_ddr4_eye_response} for the DDR4-3200 CK daisy chain interconnect. In agreement with the transient responses in Fig.~\ref{fig_ddr4_tran_response}, the predicted eye diagrams created by the different models closely match the reference eye diagrams. It is visible that the eye diagram characteristics are very similar for all four model architectures and the different training strategies. This indicates that all trained models can predict the signal behavior and eye diagram of the interconnect. There are no differences between the fully connected Dense-Dense and the LSTM-based architectures. This leads to the overall conclusion that all model types achieve comparably high prediction accuracy for scenarios with fixed design parameters.

\begin{table}[t!]
	\centering
	\caption{DDR4-3200 CK inference times on the test data ($\approx$5,000 bits)}
	\label{tab_runtimes}
	\begin{tabular}{lcccc}
		\toprule
		\textbf{Configuration} 
		& \shortstack{\textbf{Dense-}\\\textbf{Dense}} 
		& \shortstack{\textbf{Dense-}\\\textbf{LSTM}} 
		& \shortstack{\textbf{LSTM-}\\\textbf{Dense}} 
		& \shortstack{\textbf{LSTM-}\\\textbf{LSTM}} \\
		\midrule
		Sample-based TX only & 0.009 s & 0.023 s & 1.375 s & 1.383 s \\
		Sample-based TX+RX & 0.537 s & 3.357 s & 60.708 s & 62.850 s \\
		UI-based TX only & 0.004 s & 0.021 s & 0.295 s & 0.295 s \\
		UI-based TX+RX & 0.184 s & 2.871 s & 98.179 s & 100.871 s \\
		\bottomrule
		\multicolumn{5}{l}{Hardware: Nvidia RTX 4080 Super, 64 GB DDR5 RAM}\\
	\end{tabular}
\end{table}

The computational effect of these different model architectures in terms of inference runtimes is summarized in Table~\ref{tab_runtimes}. Fully connected layers achieve the lowest inference times, while the addition of LSTM layers leads to significantly higher computational cost, especially when used in the encoder. This indicates that the feedforward Dense-Dense models are more efficient for transient response prediction. The differences in inference runtime are due to the parallel sequence evaluation of dense networks, while LSTM-based models by design require sequential dependencies that slow inference. This effect becomes even stronger for the AR TX+RX recurrent implementations, because each time step is generated sequentially and then also fed back as input to the next step in a rolling forecast process, which further increases runtime.

Overall, the recurrent LSTM-based architectures result in significantly higher computational complexity and increased training and inference time due to sequential processing and gating operations. For the fixed design parameter experiments in this work, this additional complexity is not justified by the observed accuracy, because the predictive performance is very similar for all evaluated model architectures. This general behavior is likely due to the fixed design parameters of the interconnect, for which the dynamics can be learned as a quasi-static mapping from input to output. The feedforward Dense-Dense architecture in particular achieves performance comparable to the recurrent alternatives and does not require any sequential dependencies, which results in much faster training and inference. Based on the significantly lower architectural complexity as well as faster training and inference times, feedforward models are therefore the most computationally efficient approach for transient modeling in fixed design parameter spaces.

\section{Transient Modeling with Variable Design Parameters}

\begin{figure}[b!]
	\centering
	\includegraphics[width=8.5cm]{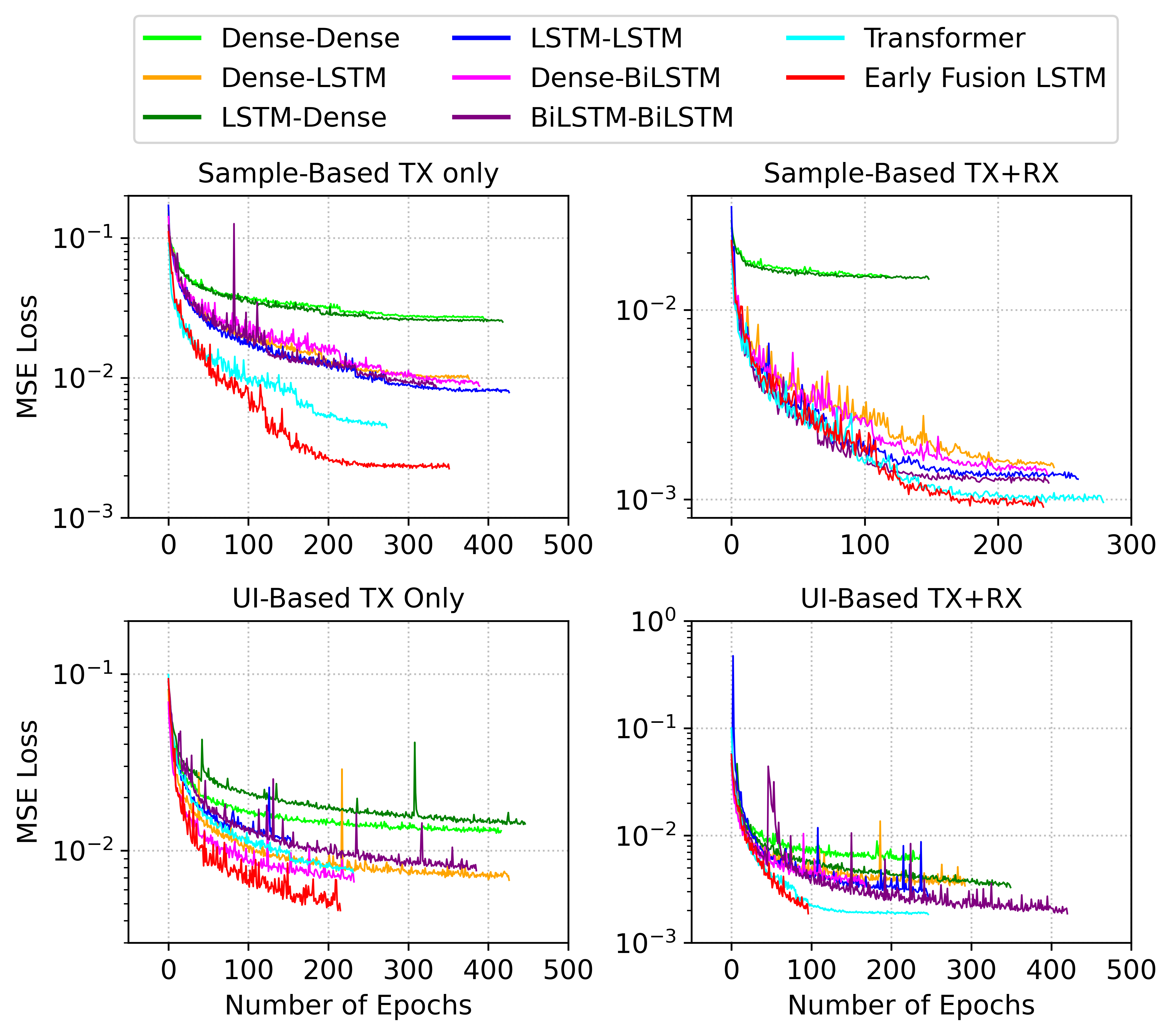}
	\caption{Validation loss comparison for all models trained on sample- and UI-based data using both TX only and RX+TX input representations.}
	\label{fig_losses}
\end{figure}

The transition from fixed design parameter scenarios to a generalized surrogate modeling approach with variable design parameters requires an architecture that can learn the dependencies between varying electromagnetic behavior and temporal signal dynamics. The 14-dimensional input design parameter vector $\mathbf{p}$ in Table~\ref{tab_ddr3_parameters} was added alongside the TX and RX sequences, which creates a high-dimensional space.

\begin{table*}[t!]
	\centering
	\caption{Variable case RMSE (mV) and $R^2$ (\%) for different training strategies and model architectures based on 100 random test channels ($\approx$100,000 bits)}
	\label{tab_variable_performance}
	\begin{tabular*}{\textwidth}{@{\extracolsep{\fill}} l *{8}{c} @{}}
		\toprule
		\multirow{2}{*}{\textbf{Model Architecture}} 
		& \multicolumn{2}{c}{\textbf{UI: TX only}} 
		& \multicolumn{2}{c}{\textbf{UI: TX+RX (AR)}} 
		& \multicolumn{2}{c}{\textbf{Sample: TX only}} 
		& \multicolumn{2}{c}{\textbf{Sample: TX+RX (AR)}} \\
		\cmidrule(lr){2-3} \cmidrule(lr){4-5} \cmidrule(lr){6-7} \cmidrule(lr){8-9}
		& \textbf{RMSE} & $\mathbf{R^2}$ & \textbf{RMSE} & $\mathbf{R^2}$ & \textbf{RMSE} & $\mathbf{R^2}$ & \textbf{RMSE} & $\mathbf{R^2}$ \\
		\midrule
		Dense-Dense & 84.3 & 98.02 & 92.3 & 97.61 & 142.1 & 94.00 & 1333.0 & -511.11 \\
		Dense-LSTM & 54.0 & 99.07 & 62.8 & 98.76 & 74.6 & 97.62 & 157.1 & 89.19 \\
		LSTM-Dense & 91.8 & 97.71 & 110.6 & 96.63 & 138.6 & 94.44 & 378.6 & 63.95 \\
		LSTM-LSTM & 77.2 & 98.31 & 68.5 & 98.60 & 66.0 & 97.74 & 231.5 & 77.70 \\
		Dense-BiLSTM & 53.3 & 99.11 & 60.8 & 98.91 & 69.4 & 97.88 & 166.7 & 79.30 \\
		BiLSTM-BiLSTM & 60.6 & 98.95 & 57.4 & 98.84 & 68.6 & 97.83 & 219.7 & 75.82 \\
		Transformer & 61.5 & 98.85 & 53.8 & 99.11 & 48.6 & 98.78 & 177.6 & 82.51 \\
		Early Fusion LSTM & 43.7 & 99.39 & 54.7 & 99.13 & 31.0 & 99.51 & 166.8 & 81.27 \\
		\bottomrule
	\end{tabular*}
\end{table*}

The training behavior in terms of validation MSE loss is shown for the evaluated architectures and training strategies in Fig.~\ref{fig_losses}. This demonstrates the differences in representational capacities of the models and shows that fully connected decoders result in the worst performance. Regardless of the complexity of the encoder, architectures with Dense decoders result in a higher loss in comparison to the other models and result in the highest prediction errors, while architectures that utilize recurrent or attention-based decoders achieve significantly lower validation losses. The losses also show a difference in architectural suitability among fusion strategies. Recurrent intermediate fusion architectures lead to improvements in comparison to the feedforward architectures, but the Early Fusion LSTM achieves the lowest losses throughout the observations. The early combination of design parameters and the temporal signal dynamics before any recurrent processing allows the Early Fusion LSTM to learn parameter-dependent signal dynamics better compared to fusing the information later in the latent space. The gap in performance is most pronounced for the sample-based TX-only configuration because the models lack the context of previous receiver signals. Therefore, they need to rely on the learned internal representation of the given channel. The UI-based losses reduce the architectural differences in performance, but still the Early Fusion LSTM achieves the lowest errors.

The transient prediction performance on the test dataset is summarized in Table~\ref{tab_variable_performance} and confirms the trends observed with respect to the loss behavior of Fig.~\ref{fig_losses}. Models with fully connected decoders lead to the highest errors for all training regimes due to their inability to predict complex temporal sequences, while intermediate latent fusion with recurrent decoders, such as the LSTM-LSTM and Dense-BiLSTM, leads to clear improvements but stays suboptimal. Transformer architectures are performance-wise located in between the recurrent intermediate fusion and the most accurate Early Fusion LSTM architecture. The sample-based TX-only Early Fusion LSTM is the most accurate surrogate model overall with an RMSE of 31.0~mV and an $R^2$ of 99.51\% for transient prediction. This demonstrates that early parameter injection without relying on previous RX states results in the highest transient prediction accuracy, as error accumulation is not a concern in this case. In contrast, the AR TX+RX sample-based training strategy leads to instability due to exactly that sequential error accumulation. The fully connected Dense-Dense architecture is heavily struck by this error accumulation in combination with its inability to predict complex temporal output signals, which results in an $R^2$ of -511.11\%, but also the more complex models show subpar transient prediction accuracy. The UI-based segmentation stabilizes training and aligns each UI-based prediction, which makes the models more robust against error propagation. This is consistent with the loss behavior analysis. Still, the same structural limitations with slightly reduced margins between architectures can be observed, as feedforward decoders lag behind the recurrent decoders and the Early Fusion LSTM remains the superior architecture for achieving the highest accuracy for the UI-based approach as well.

\begin{figure}[t!]
	\centering
	\includegraphics[width=8.5cm]{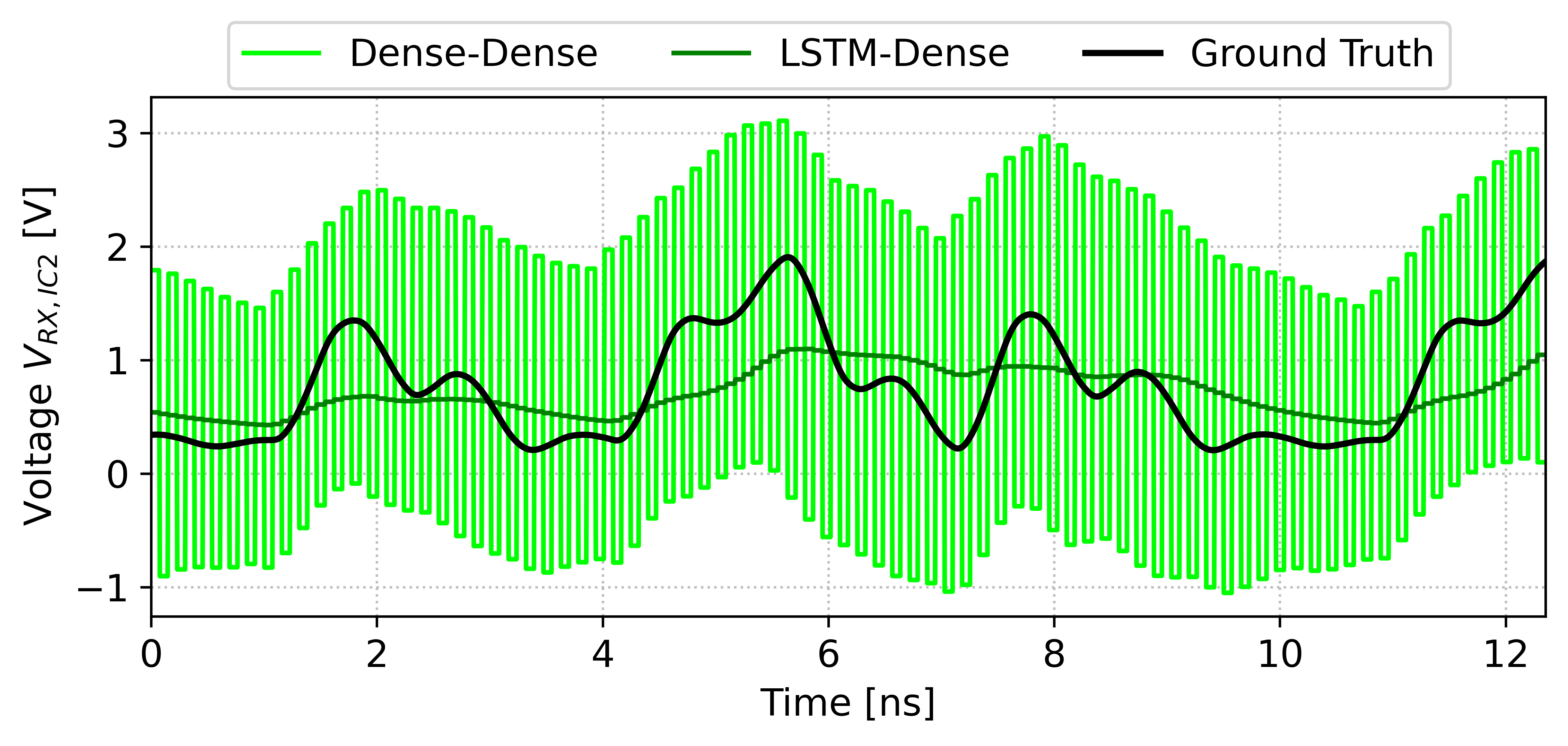}
	\caption{Transient response prediction example using the TX+RX AR sample-based training strategy and intermediate fusion fully connected decoder architectures.}
	\label{fig_tran_tx+rx_failure}
\end{figure}

\begin{figure}[b!]
	\centering
	\includegraphics[width=8.5cm]{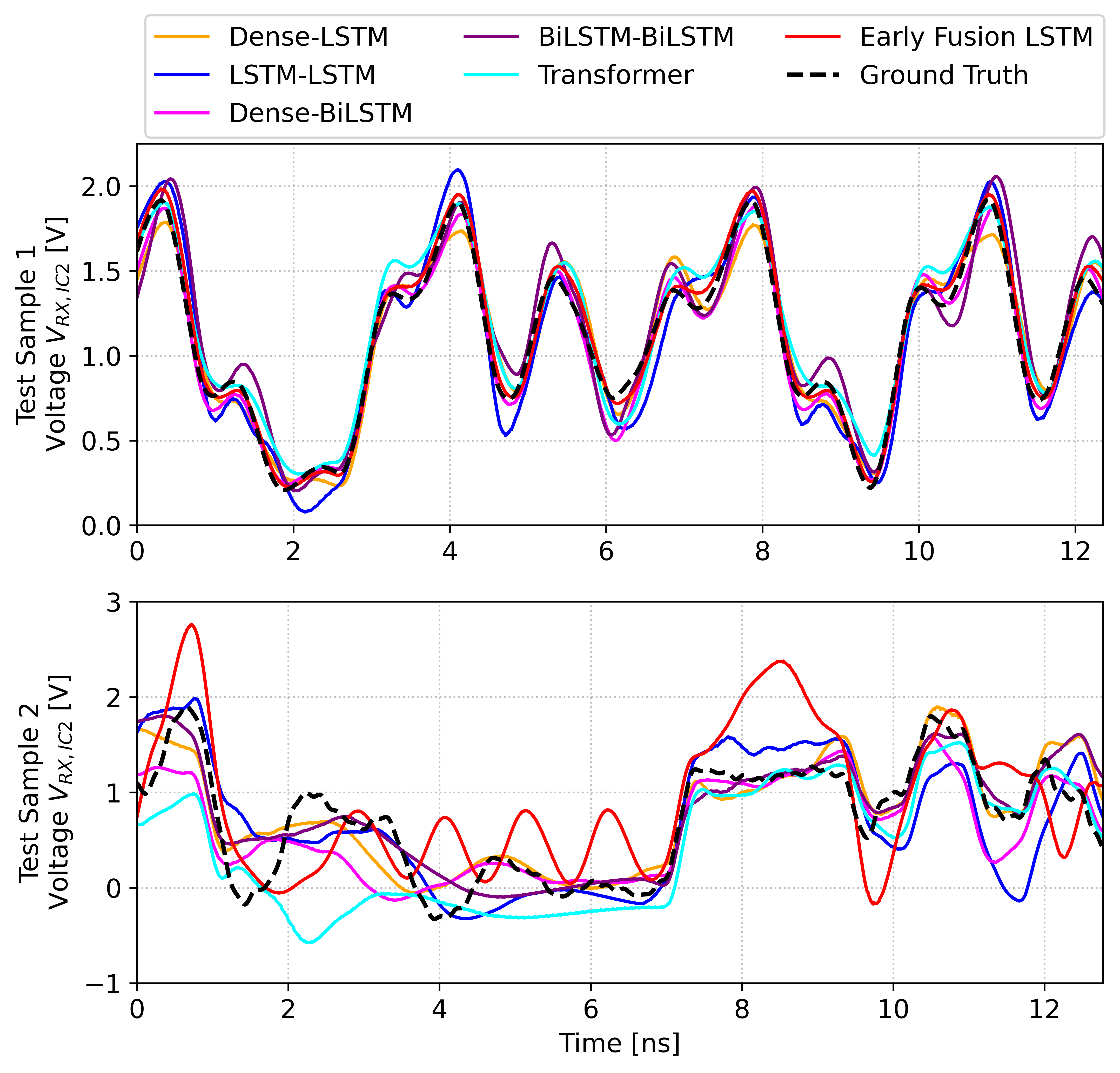}
	\caption{Transient response prediction examples using the TX+RX AR sample-based training strategy for intermediate fusion recurrent decoder, Transformer and Early Fusion LSTM architectures.}
	\label{fig_tran_input+output}
\end{figure}

To also interpret the performance metrics from Table~\ref{tab_variable_performance} visually, the transient waveform predictions are analyzed for the different training strategies, which confirms the limitations of AR prediction and the overall superiority of the early fusion LSTM architecture for transient modeling in variable design scenarios.

\begin{figure}[t!]
	\centering
	\includegraphics[width=8.5cm]{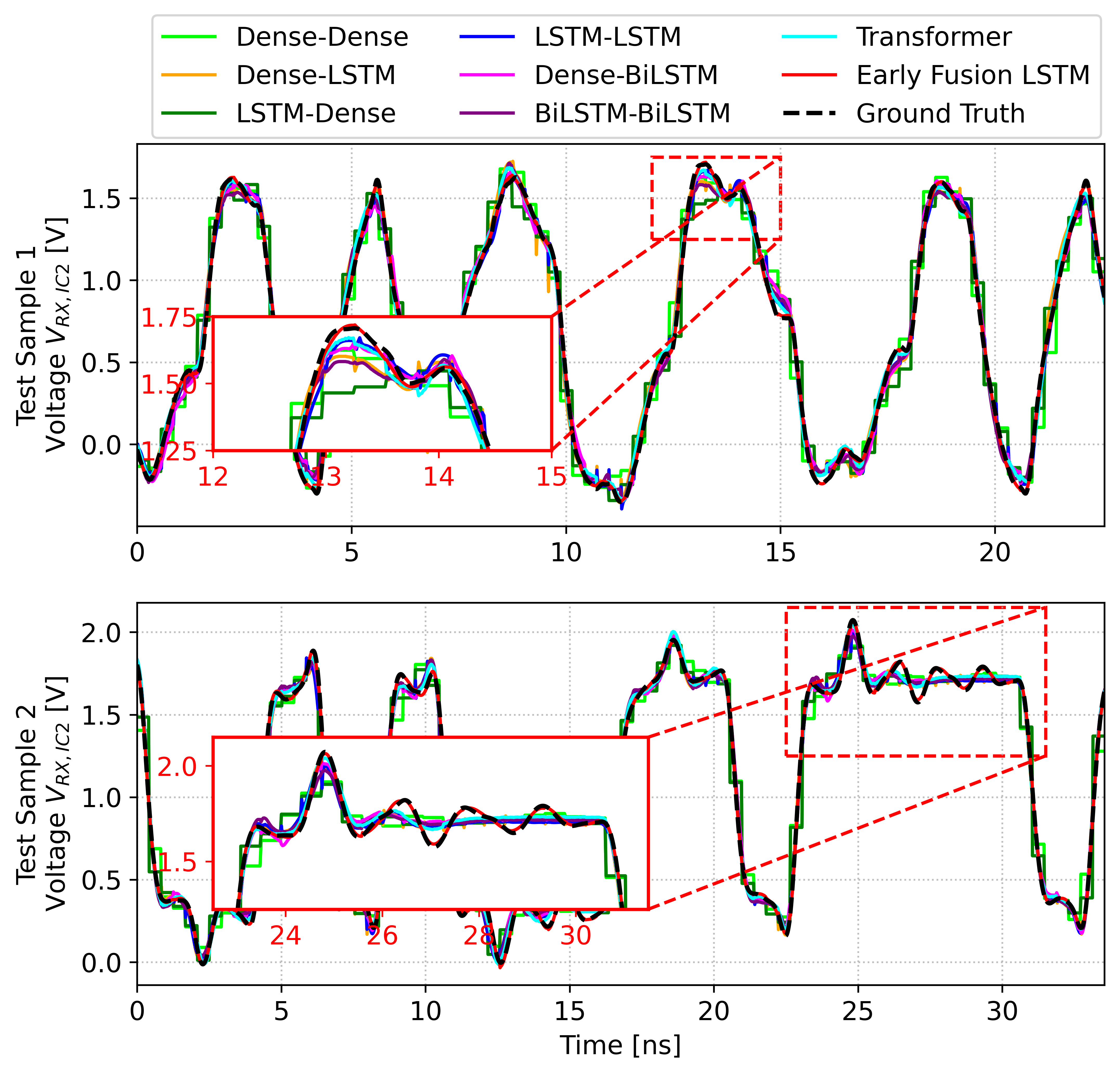}
	\caption{Transient response prediction examples using the TX-only sample-based training strategy.}
	\label{fig_tran_tx_only}
\end{figure}

The failure of feedforward decoders in the sample-based TX+RX AR regime is also visible in Fig.~\ref{fig_tran_tx+rx_failure}. The accumulation of error causes the Dense-Dense architecture to oscillate wildly, which demonstrates its numerical instability. Similarly, the LSTM-Dense model completely fails to predict the signal dynamics accurately, with an attenuated representation of the waveform and moving-average characteristics. As shown in Fig.~\ref{fig_tran_input+output}, this AR instability is not limited to feedforward decoders. Higher-capacity recurrent models provide reasonable transient response predictions for specific design parameter constellations, e.g., in Test Sample 1, but when exposed to a different parametric variation, e.g., in Test Sample 2, all evaluated models in the AR regime fail to make accurate predictions and suffer from phase drift and amplitude deviation. This confirms that AR and sequential receiver feedback are unreliable for variable design parameter scenarios in this comparison. This occurs because variable design parameters impact the channel dynamics, and a minor prediction error may be sufficient to push the autoregressive feedback into an unseen area, which causes the AR rollout to progressively decouple from the ground-truth physical response.

\begin{table*}[t!]
	\centering
	\caption{Variable case eye diagram metrics: eye-height (EH) [mV] and eye-width (EW) [ps] prediction accuracy in terms of RMSE (mV) and $R^2$ (\%) for different training strategies and model architectures based on 100 random test channels ($\approx$100,000 bits)}
	\label{tab_variable_ehew}
	\begin{tabular*}{\textwidth}{@{\extracolsep{\fill}} l *{8}{c} @{}}
		\toprule
		\multirow{2}{*}{\textbf{Model Architecture}} 
		& \multicolumn{2}{c}{\textbf{UI: TX only}} 
		& \multicolumn{2}{c}{\textbf{UI: TX+RX (AR)}} 
		& \multicolumn{2}{c}{\textbf{Sample: TX only}} 
		& \multicolumn{2}{c}{\textbf{Sample: TX+RX (AR)}} \\
		\cmidrule(lr){2-3} \cmidrule(lr){4-5} \cmidrule(lr){6-7} \cmidrule(lr){8-9}
		& \textbf{RMSE} & $\mathbf{R^2}$ \textbf{(\%)} 
		& \textbf{RMSE} & $\mathbf{R^2}$ \textbf{(\%)} 
		& \textbf{RMSE} & $\mathbf{R^2}$ \textbf{(\%)} 
		& \textbf{RMSE} & $\mathbf{R^2}$ \textbf{(\%)} \\
		\midrule
		\multicolumn{9}{c}{\textbf{Eye-Height (EH) [mV]:}} \\
		\midrule
		Dense-Dense & 78.8 & 95.45 & 102.7 & 92.28 & 168.6 & 87.12 & 464.0 & -8.92 \\
		Dense-LSTM & 53.3 & 97.92 & 60.2 & 97.35 & 98.0 & 95.65 & 132.5 & 91.11 \\
		LSTM-Dense & 75.9 & 95.78 & 126.3 & 88.33 & 157.1 & 88.81 & 459.7 & -6.92 \\
		LSTM-LSTM & 67.8 & 96.63 & 78.1 & 95.54 & 79.1 & 97.83 & 204.1 & 78.63 \\
		Dense-BiLSTM & 48.0 & 98.31 & 54.5 & 97.83 & 80.2 & 97.03 & 195.9 & 80.58 \\
		BiLSTM-BiLSTM & 66.3 & 96.78 & 53.2 & 97.93 & 89.1 & 96.41 & 322.3 & 47.44 \\
		Transformer & 64.0 & 97.00 & 58.8 & 97.47 & 52.7 & 98.74 & 209.3 & 77.83 \\
		Early Fusion LSTM & 34.5 & 99.13 & 49.8 & 98.19 & 30.7 & 99.57 & 148.8 & 88.79 \\
		\midrule
		\multicolumn{9}{c}{\textbf{Eye-Width (EW) [ps]:}} \\
		\midrule
		Dense-Dense & 113.2 & 83.11 & 106.6 & 84.74 & 257.5 & 88.82 & 588.4 & 40.37 \\
		Dense-LSTM & 66.6 & 94.16 & 91.9 & 88.74 & 194.1 & 93.65 & 130.0 & 97.07 \\
		LSTM-Dense & 91.9 & 88.87 & 114.4 & 82.56 & 240.4 & 90.25 & 379.4 & 75.25 \\
		LSTM-LSTM & 109.9 & 84.09 & 88.0 & 89.67 & 113.0 & 97.85 & 214.8 & 92.05 \\
		Dense-BiLSTM & 68.9 & 93.75 & 52.9 & 96.27 & 201.0 & 93.19 & 186.9 & 94.00 \\
		BiLSTM-BiLSTM & 67.3 & 94.04 & 58.4 & 95.46 & 161.4 & 95.61 & 200.2 & 93.05 \\
		Transformer & 116.3 & 82.19 & 50.1 & 96.65 & 52.6 & 99.53 & 211.2 & 92.32 \\
		Early Fusion LSTM & 35.3 & 98.35 & 30.3 & 98.78 & 19.5 & 99.94 & 89.6 & 98.62 \\
		\bottomrule
	\end{tabular*}
\end{table*}

In comparison, the sample-based TX-only predictions illustrated in Fig.~\ref{fig_tran_tx_only} demonstrate the previously observed strong representational gap between the architectures. Without the crutch of AR feedback, models with dense decoders output averaged and staircase-like approximations that completely fail to produce the fine transient response details. This is due to the fully connected layers that attempt to map a static latent vector directly to a temporal output, which makes it difficult to predict the waveform details conditioned on the design parameters. On the other hand, the Early Fusion LSTM shows the most accurate transient predictions. In both test samples, the Early Fusion LSTM model tracks the ground truth almost perfectly, including fine transient details such as voltage overshoots and ringing effects shown in the zoomed regions. The Transformer and recurrent intermediate fusion architectures do not show the dense decoder staircase behavior, but they also fail to predict the temporal details and overshoot effects that only the Early Fusion LSTM model is able to predict accurately.

\begin{figure}[b!]
	\centering
	\includegraphics[width=8.5cm]{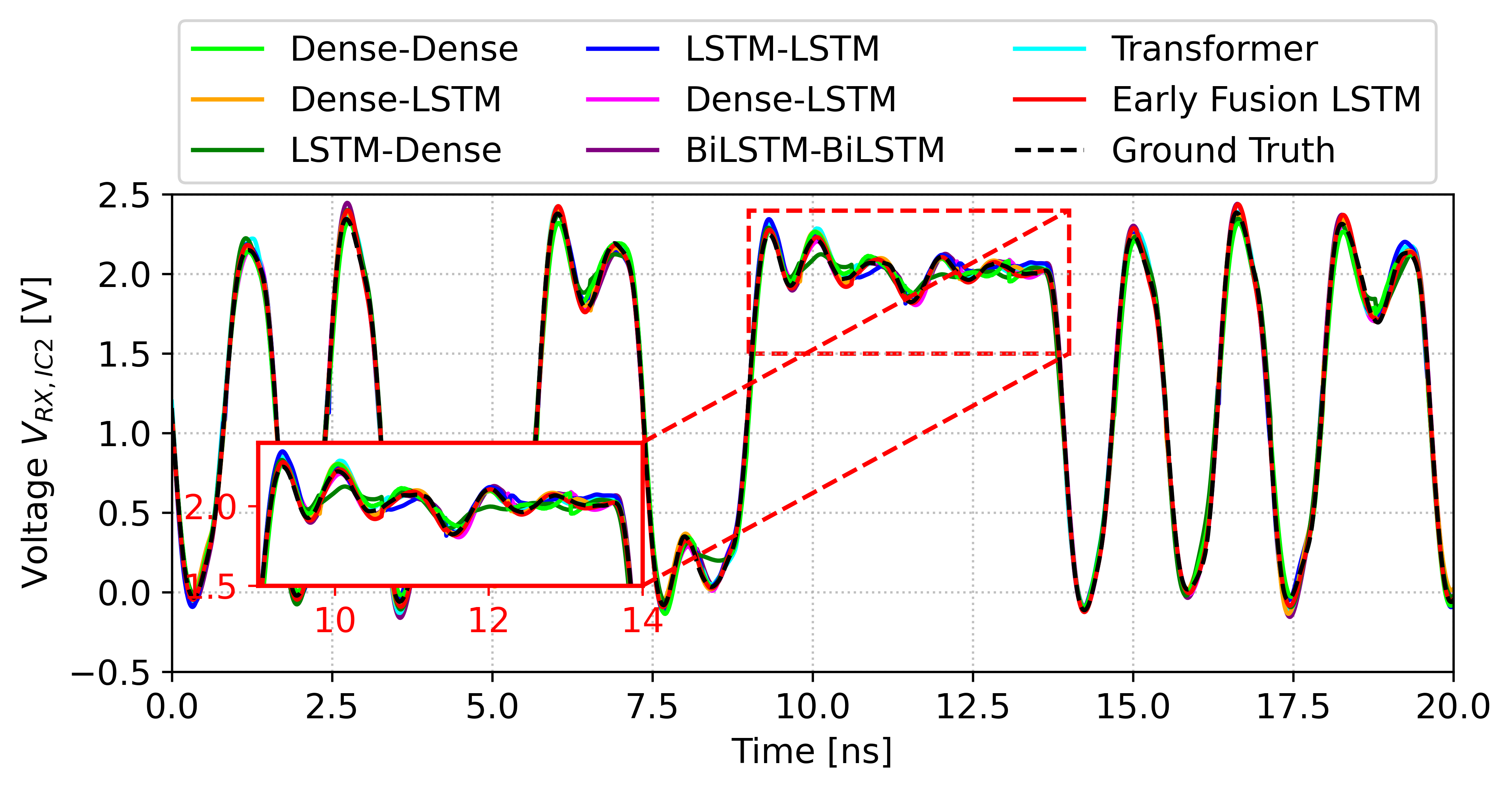}
	\caption{Transient response prediction example using the TX-only UI-based training strategy.}
	\label{fig_ui_tx_only}
\end{figure}

\begin{figure}[b!]
	\centering
	\includegraphics[width=8.5cm]{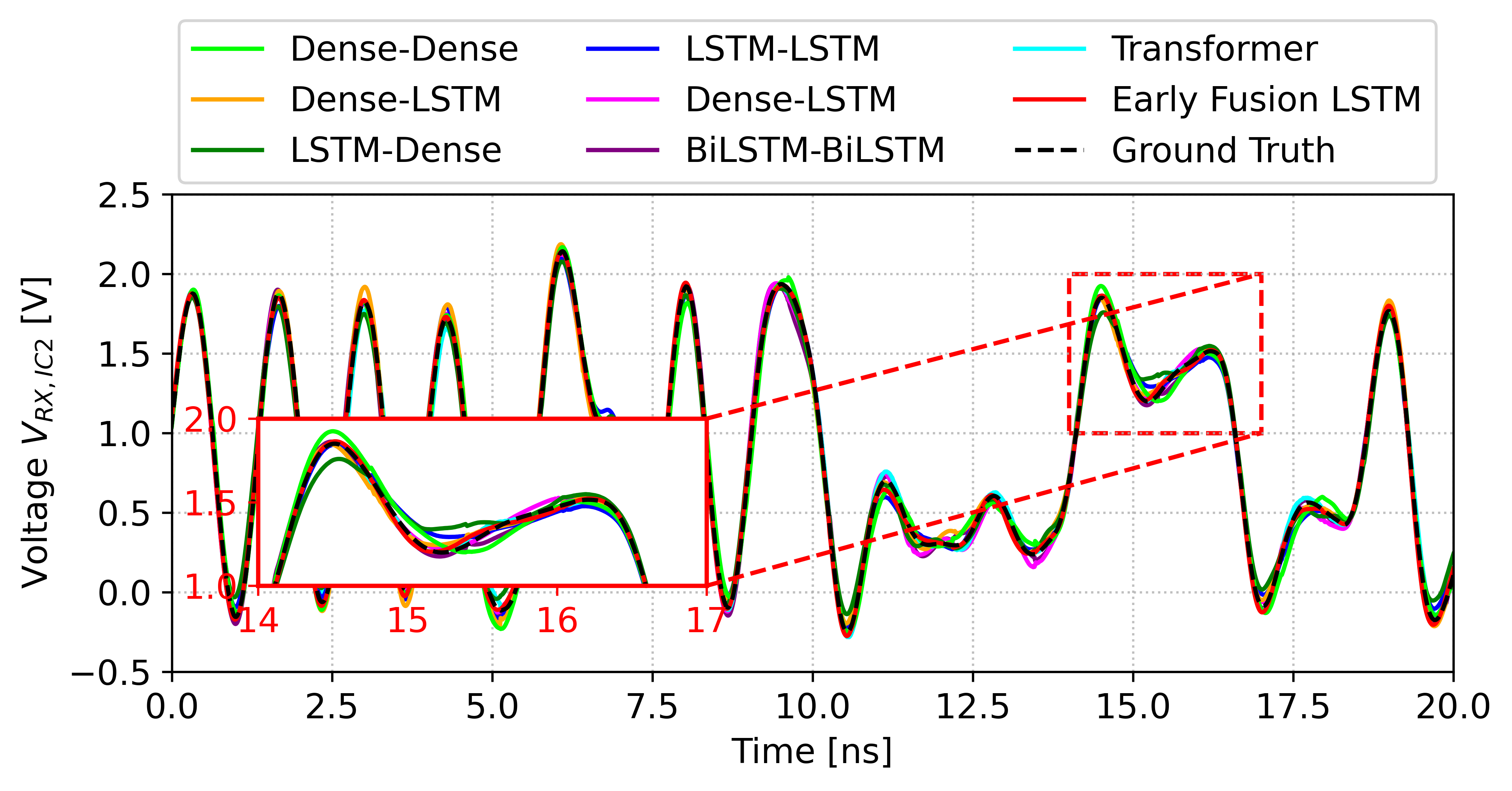}
	\caption{Transient response prediction examples using the TX+RX AR UI-based training strategy.}
	\label{fig_ui_tx+rx}
\end{figure}

Finally, the UI-based predictions for both the TX-only and TX+RX AR strategies, shown in Fig.~\ref{fig_ui_tx_only} and Fig.~\ref{fig_ui_tx+rx}, respectively, demonstrate more consistent performance for all architectures. As the models now predict localized UI segments rather than random sample-based sequences, the UI-based formulation acts as an alignment method that prevents the error accumulation observed for the sample-based AR models. However, as can be seen in the zoomed regions in both figures, the models show minor prediction inaccuracies when predicting the overshoot regions. The UI-based strategy is much more robust through the structural alignment of the transient sequence. Overall, transient prediction performance is stabilized through the UI-based strategy, but at the same time this also prevents the models from reaching the maximum accuracy, which the Early Fusion LSTM achieves in the sample-based TX-only case.

\begin{figure}[t!]
	\centering
	\includegraphics[width=8.5cm]{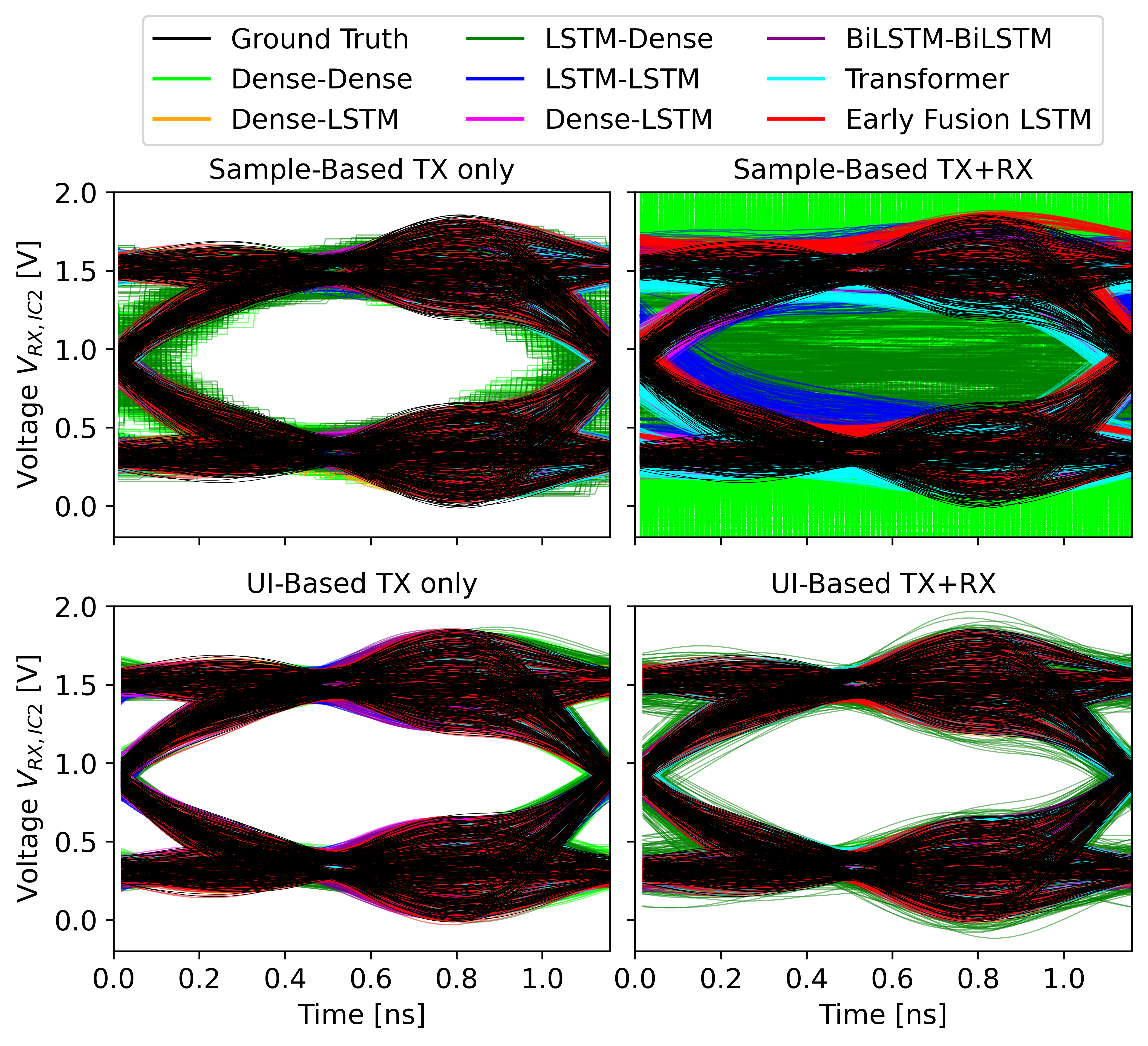}
	\caption{Eye prediction examples for the variable case with different model architectures and training strategies.}
	\label{fig_eye_variable}
\end{figure}

To evaluate the practical implications for SI analysis, the transient predictions are used to create eye diagrams, as illustrated in Fig.~\ref{fig_eye_variable}, including the corresponding statistical metrics for eye-height (EH) and eye-width (EW) summarized in Table~\ref{tab_variable_ehew}. The visual representation in Fig.~\ref{fig_eye_variable} aligns with the prior observations with respect to the transient waveform prediction. In the sample-based TX-only regime, the Early Fusion LSTM overlays the ground truth very accurately, whereas feedforward decoder architectures show severe inner-eye closure and blocky approximations due to the transient staircase behavior. This visual accuracy is also validated by the EH/EW metrics in Table~\ref{tab_variable_ehew}, as the Early Fusion LSTM achieves the most accurate EH and EW metrics overall. In contrast, the instability of the sample-based AR (TX+RX) approach corrupts the eye metrics as expected, resulting in a negative EH $R^2$ of -8.92\% for the Dense-Dense model. However, even higher-capacity models show significant EH/EW errors compared to the TX-only case, which confirms that RX feedback and AR are unsuitable to generate accurate eye diagrams. Finally, while the UI-based strategies successfully act as an alignment method to prevent complete eye closure and keep the $R^2$ values positive for all models, they decrease overall prediction performance, which is reflected in the EH and EW metrics. Overall, these eye diagram metrics validate that sample-based TX-only sequence modeling combined with early parameter fusion provides the highest predictive accuracy for both transient modeling and eye diagram prediction in variable design parameter scenarios.

Moreover, these results are particularly significant when compared to the direct regression approach of \cite{Withoeft2025}, which maps design parameters $\mathbf{p}$ to EH and EW using a fully connected neural network. Using the identical data distribution, that method results in an EW $R^2$ of 95\% at 165~ps RMSE and an EH $R^2$ of 82\% at 194~mV RMSE. This strong difference in performance confirms that seq2seq transient modeling results in higher accuracy because it models the underlying temporal electromagnetic behavior of the high-speed link. In contrast to black-box parameter-to-metric surrogates, modeling the full transient response allows the network to learn the underlying dynamic signal behaviors, which results in significant accuracy gains and also provides transparency with respect to the exact waveform responsible for the predicted eye metrics.

\section{Conclusion and Future Work}

This paper presents sequence-to-sequence neural surrogate models for high-speed signal integrity analysis in fixed and variable design parameter spaces. The results demonstrate that the optimal surrogate architecture is determined by the dimensionality and variability of the design space. For fixed design parameter cases, feedforward architectures are sufficient as they deliver high predictive accuracy and maximize computational efficiency. However, modeling variable design spaces requires sequence-aware architectures to model the dynamic dependencies between the electromagnetic physics of the conditioning design parameters and the transient signal behavior. In this scenario, the non-autoregressive sample-based Early Fusion LSTM is the superior architecture. Injection of design parameters into the initial encoding stage combined with non-autoregressive training allows the model to avoid error accumulation seen in autoregressive modeling and results in superior accuracy in both transient waveform prediction, with 31.0~mV RMSE and 99.51\% $R^2$, as well as in eye-diagram metrics, where it achieves an eye-height prediction of 30.7~mV RMSE at an $R^2$ of 99.57\% and an eye-width prediction of 19.5~ps RMSE at an $R^2$ of 99.94\%.

Future work will extend this sequence-to-sequence methodology to more complex interconnect structures and effects. This includes the application to actively equalized high-speed links, such as SerDes channels utilizing FFE, CTLE, and DFE. Parallel research will focus on the optimization of computational efficiency of the recurrent non-autoregressive early fusion architectures to accelerate large-scale design space exploration for transient response modeling with variable design parameters. This will include further investigation of lightweight SRU/GRU recurrent variants in combination with optimization of structural hyperparameters, such as minimizing hidden dimensions, layer depth, and the required TX lookback window, for even higher computational efficiency while preserving the strong transient prediction accuracy demonstrated in this work.

\bibliography{paper_bib}

\bibliographystyle{IEEEtran}

\end{document}